# Synergistic effects of ferromagnetic elements and LAGP solid electrolyte in suppressing and trapping polysulfide shuttle transfers in lithium–sulfur batteries

Giovanni Ceccio*, Jiri. Vacík, Mykhailo Drozdenko, Romana Mikšová, Josef Novak, Eva Štěpanovská, Mayur Khan

Department of Neutron and Ion Methods, Nuclear Physics Institute (NPI) of the Czech Academy of Sciences, 25068 Husinec-Řež, Czech Republic

Corresponding author: ceccio@ujf.cas.cz

**Abstract**

The large-scale commercialization of promising lithium–sulfur (Li–S) batteries remains limited by the polysulfide shuttle effect, which causes rapid capacity fading and poor cycle life. In this study, we present a scalable strategy to mitigate this challenge by modifying polyethylene (PE) separators with ferromagnetic and solid-state ionic coatings. Thin films of nickel (Ni), cobalt (Co), and the Li-ion–conducting ceramic $Li_{1.5}Al_{0.5}Ge_{1.5}(PO_4)_3$ (LAGP) were deposited via ion beam sputtering, while Ni ion implantation was also employed to modify the PE substrate. The electrochemical performance of pristine and modified separators was evaluated using electrochemical impedance spectroscopy (EIS) and staircase voltammetry (SV) in liquid electrolyte within H-cell configurations. Surface morphology and elemental composition were characterized by scanning electron microscopy (SEM) and Rutherford backscattering spectroscopy (RBS). The results show that LAGP-based coatings significantly enhance separator stability and effectively suppress polysulfide diffusion, leading to lower redox peak intensities and improved cycling performance. In contrast, Ni coatings exhibited poor long-term stability, likely due to parasitic reactions or delamination during its life time. The combined LAGP/Co architecture provided the most effective suppression of the polysulfide shuttle, attributed to synergistic ionic and catalytic effects that promote interfacial stability and selective ion transport. Ni implantation into PE showed only a negligible effect. This study highlights the potential of integrating solid ionic conductors with ferromagnetic layers to design multifunctional separators for high-performance Li–S batteries.

## 1. Introduction

Lithium-sulfur (Li-S) batteries are a promising alternative to lithium-ion batteries for next-generation rechargeable energy storage, owing to their exceptionally high capacity, low cost, natural abundance, and the reduced environmental impact of sulfur compared with Li-ion systems [1-4]. With a high theoretical specific capacity of 1675 mAh g$^{-1}$ and an energy density of 2600 Wh kg$^{-1}$ [5], Li-S batteries are also considered strong candidates for the development of ultrathin all-solid-state batteries, where high energy density is a crucial parameter to compensate for reduced device dimensions. Despite these advantages, the commercialization of Li-S batteries remains hindered by several challenges, most notably their limited cycle life. This poor cycling stability primarily arises from the high solubility of lithium polysulfides generated during the charge/discharge processes. The polysulfide species can migrate through the separator and shuttle between the electrodes, leading to active material loss, reduced Coulombic efficiency, and rapid capacity fading.

To overcome this bottleneck and suppress the shuttle effect, several strategies have been proposed and extensively investigated [6]. For instance, incorporating electrocatalytic materials such as carbon nanostructures into the cathode matrix has been explored to enhance redox kinetics and increase sulfur loading [7-9]. Another approach involves the use of transition metals, which can chemically adsorb lithium polysulfides through strong chemisorption interactions [10-12]. A further strategy focuses on introducing sophisticated physical barriers, particularly through the modification of separators. These separators are essential components in Li-S batteries, functioning not only as ionic conductors but also as physical barriers that can mitigate polysulfide migration. One promising route to enhance this barrier effect involves coating the separator surface with a thin layer of material capable of trapping just polysulfides on the cathode side via physical adsorption (or chemical bonding), thereby inhibiting their diffusion and so reducing the shuttle effect.

In this work, we propose a scalable method for modifying separators using ultra-thin films or ion implantation to suppress the shuttle effect while simultaneously enhancing the ionic conductivity of the separator. Previous studies have demonstrated that certain metals can reduce the shuttle effect and improve cycling stability when applied as coatings on current collectors [13]. In our experiments, commercially available polyethylene (PE) separators were modified using metal coating and ion implantation techniques. To preserve high Li-ion conductivity, the solid-state ionic conductor $Li_{1.5}Al_{0.5}Ge_{1.5}(PO_4)_3$ (LAGP), a LISICON-type material, was employed as a coating layer on the separator. LAGP is a Li-ion–conducting ceramic with a NASICON (Na superionic conductor)–type structure, and its influence on separator performance was investigated both with and without additional nickel (Ni) or cobalt (Co) coatings deposited by ion beam sputtering. Furthermore, a complementary approach was explored using Ni-implanted PE separators prepared via ion implantation at the Tandetron accelerator (NPI CANAM infrastructure, [14]). All modified separators were subsequently tested in a liquid electrolyte using an H-cell configuration and subjected to electrochemical cycling. The amount of sulfur trapped within the separator after cycling was quantified by Rutherford Backscattering Spectroscopy (RBS).

## 2. Experimental section

### 2.1 Materials

Lithium bis(trifluoromethanesulfonyl)imide (LiTFSI, Thermo Fisher Scientific, [15]) was dissolved in a mixed solvent of 1,3-dioxolane (DOL) and 1,2-dimethoxyethane (DME) in a 1:1 volume ratio (both solvents purchased from Thermo Fisher Scientific). Carbon-coated electrodes (MTI Corporation, [16]) and sulfur electrodes (Merck Life Science, [17]) were cut into identical pieces with an area of 625 mm² and used in H-cell configurations to evaluate the performance of the modified PE separators. As separator, polyethylene (PE) foils (Nanografi, [18]) were applied, partially fluorinated (enhancing properties [19]). The PE separators were modified by sputter deposition of thin films using high-purity targets of nickel (J.K. Lesker, [20]), cobalt (J.K. Lesker), and LAGP (MSE Supplies, [21]). The LAGP solid electrolyte was selected because, among the available NASICON-type ceramic phosphate electrolytes on the market, it exhibits the highest ionic conductivity [22].

## 2.2 PE separator modification

A commercially available polyethylene (PE) separator, 12 μm in thickness (Nanografi), was systematically modified through ion-beam implantation and thin-film coating techniques. Ion implantation was carried out using a 3 MV Tandetron 4130 accelerator [14] with a 1.1 MeV Ni ion beam. The separators were implanted with ion fluences of $5 \times 10^{12}$, $1 \times 10^{13}$, and $5 \times 10^{13}$ Ni cm$^{-2}$, the middle fluence corresponds to the threshold of ion-track overlapping and marks the onset of radiation damage within the PE polymer matrix [23] (PE can tolerate absorbed dose on the order of tens of kGy promoting both crosslinking and chain scission defects). At the selected Ni ion energy, the projected ion range was ~ 1.52 μm, as estimated by SRIM simulation software [24]. Thus, the PE separators were modified only within the near-surface region of the polymer matrix, according to the requirements to form a thin Ni barier, which would function as a permeable layer allowing the passage of Li$^+$ ions on one side and the capture of complex polysulfide ions on the other side.

For the second modification approach, surface coating of the separators with feroic metals was carried out using the ion-beam sputtering (IBS) technique. Ultrathin coatings were deposited at the Low Energy Ion Facility (LEIF, laboratory-developed NPI deposition/irradiation system [14]). High-purity metallic targets of Ni (4 N purity, J. K. Lesker, 2" × 0.125") or Co (4 N purity, J. K. Lesker, 2" × 0.125") were employed in the IBS. For the subsequeent LAGP deposition, a 16 mm-diameter ceramic sample (MSE Supplies) was used as the sputtering target. All materials were placed in a center of an ultrahigh-vacuum (UHV) chamber and sputtered by 20 keV Ar$^+$ ion beam (with a current of ~ 100 μA), generated by a Duoplasmatron ion source (laboratory-built, NPI). The sputtered species (Ni/Co atoms and LAGP clusters) were deposited onto the PE substrates positioned 120 mm from the target (see Fig. 1). Preliminary deposition tests were conducted with varying sputtering times (10 minutes – 1 hour) to determine optimal conditions prior to fabricating the final multilayer structures consisting of the LAGP and metallic films (PE/LAGP/Ni, PE/LAGP/Co).

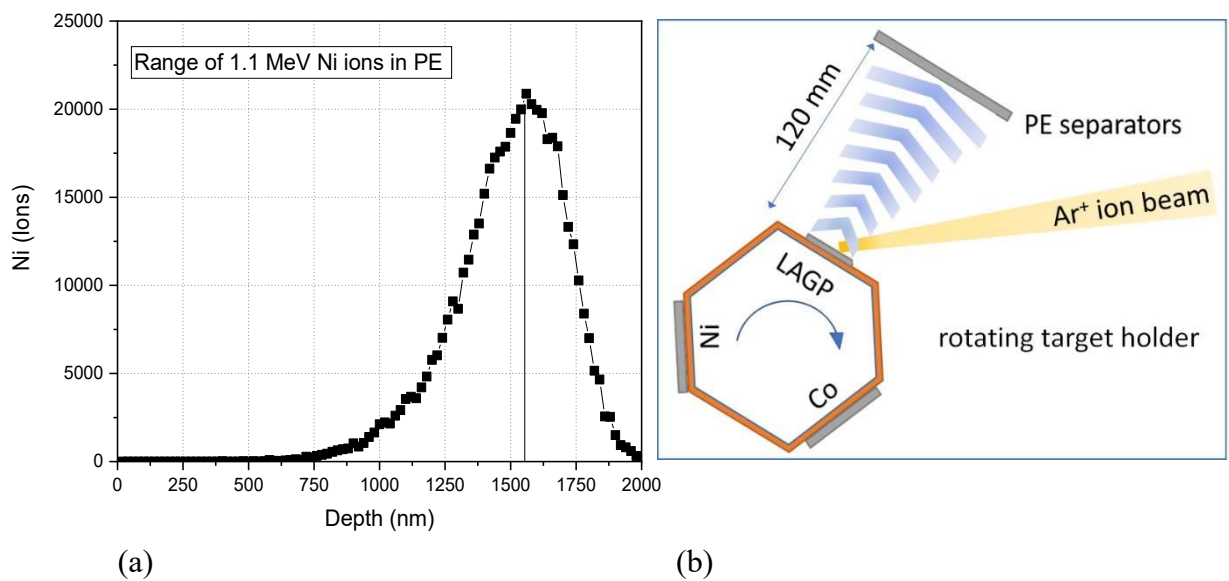

(a)      (b)

**Fig. 1.** **(a)** Range of 1.1 MeV Ni ions in polyethylene PE; **(b)** principle setup of the IBS technique.

## 2.3 Electrochemical testing of separators in a H-cell

To evaluate the sulfur trapped in the modified separators, PE foils with implanted nickel or LAGP/Ni (Co) coatings were immersed and cycled for an equal duration time (1 hour) into a H-cell (lab-made using Teflon) and cyclically tested in potentiostatic mode under identical conditions. Symmetric cells equipped with platinum electrodes were used throughout the experiment. After testing, the samples were rinsed with deionized water for 10 minutes and subsequently dried at RT for 24 hours in a controlled ambient atmosphere.

## 2.4 Characterization

The modified separators were after electrochemical testing examined using scanning electron microscopy (SEM, TM4000 Plus, Hitachi High-Tech Europe GmbH, Germany), to analyze their surface morphology, and by Rutherford Backscattering Spectroscopy (RBS), to determine the elemental composition and profiling (including thickness) of the deposited films. For the RBS measurements were used 2.0 MeV $He^+$ ions with a 1 $mm^2$ beam size and a current of ~ 5 nA, generated at the Tandetron accelerator [14]. The backscattered ions were measured using an Ultra-Ortec PIPS detector with a 5.4 mm circular aperture positioned ~ 70 mm from the sample holder at a scattering angle of 170°. The energy resolution of the RBS system was ~ 20 keV. The obtained energy spectra were evaluated using the SIMNRA software package [25].

## Results

Initial insights into the coating morphology were obtained from SEM analyses of Ni-coated PE separators prepared with deposition times of 20, 40, and 60 minutes. The corresponding SEM micrographs are presented in Fig. 2. As shown in the images, the characteristic fibrous texture of the polyethylene separator remains discernible even after Ni deposition, indicating that the nickel layer does not form a fully continuous film at any of the investigated deposition times. Instead, it preserves the porous architecture of the polymer substrate. This behavior is consistent with the intended purpose of the IBS process to produce a thin, yet porous Ni overlayer while maintaining the intrinsic permeability of the separator, a parameter essential for sustaining efficient ion transport in electrochemical systems.

At shorter deposition times, the Ni coating remains largely discontinuous, manifested as micro-agglomerates distributed across the PE surface. These agglomerates likely arise from the nucleation and subsequent growth of Ni nanoparticles during the early stages of film formation [26]. As the deposition time increases, the surface coverage gradually improves, ultimately giving rise to a more coherent, fibrous-like coating morphology – particularly pronounced after one hour of sputtering.

As noted above, Ni plays a crucial role in sulfur capture, which is particularly important for the performance of lithium-sulfur batteries, where suppressing polysulfide migration is essential for achieving improved electrochemical stability. The depth distribution of Ni and the extent of sulfur capture on or beneath the surface of the PE separators were analyzed using RBS. Fig. 3a presents the spectra obtained for pristine separators and those implanted with Ni at fluences of $10^{12}$ $cm^{-2}$, $10^{13}$ $cm^{-2}$, and 5 x $10^{13}$ $cm^{-2}$. As evident from the data, the signal levels in the regions corresponding to expected S and Ni energies are very low for all fluences (showing poor statistics, only up to several counts per channel) and do not differ significantly from those of the pristine PE foil. Moreover, a portion of the background in the region between approximately channels 200 and 400 is likely attributable to pile-up events, which introduces additional uncertainty into the spectral evaluation. Fig. 3b shows the evaluated integrated areas of the regions assigned to S and Ni for all selected Ni probe fluences (including the unirradiated foil).

Due to the limited counting statistics, the extracted areas are associated with substantial uncertainties; the error bars in the figure just represent the standard deviations

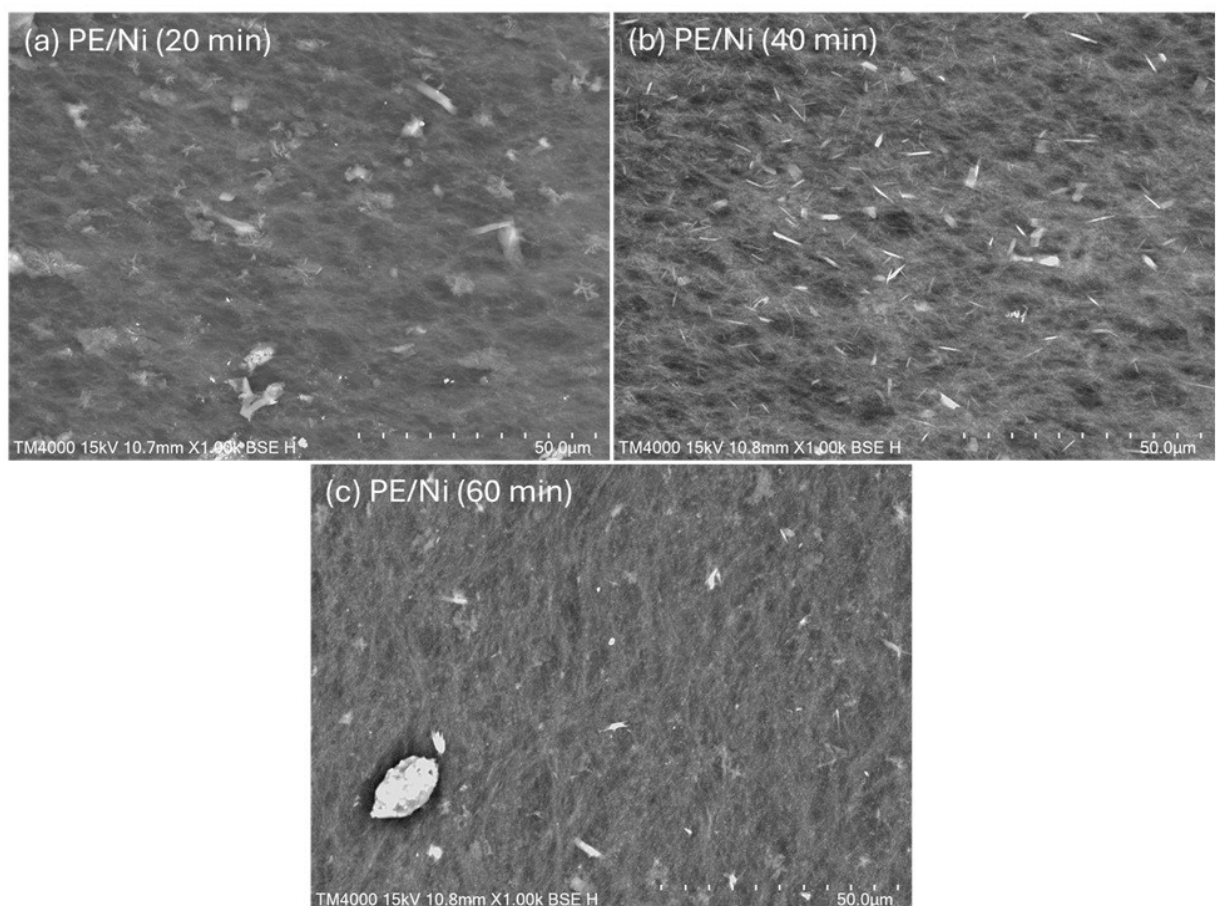

**Fig. 2.** SEM micrographs of Ni-coated PE separators prepared by IBS at deposition times of **(a)** 20 min, **(b)** 40 min, and **(c)** 60 min.

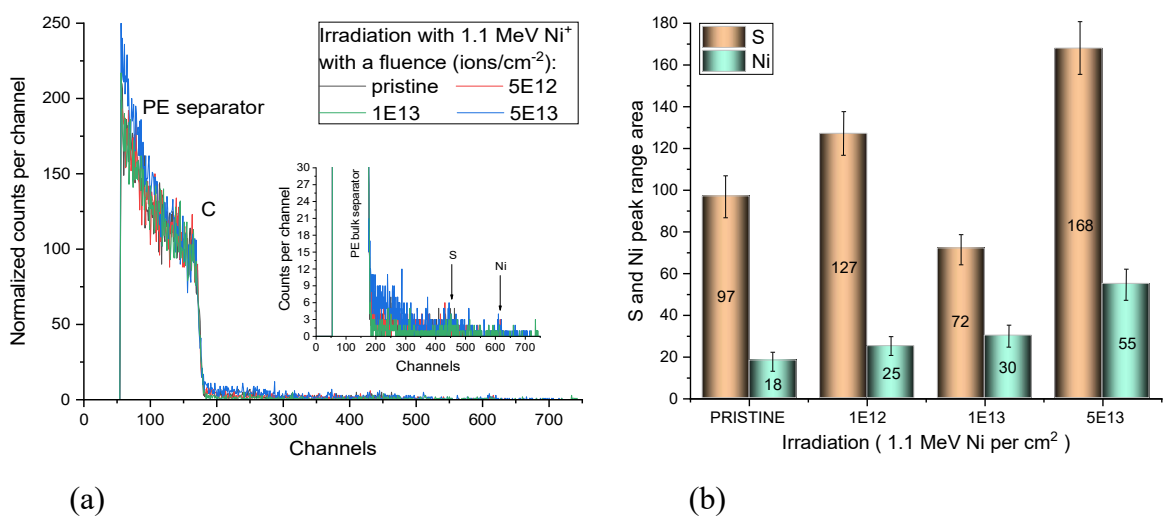

**Fig. 3. (a)** RBS spectra of PE separators implanted with 1.1 MeV Ni ions at fluences of 1 x $10^{12}$, 1 x $10^{13}$, and 5 x $10^{13}$ at. cm$^{-2}$, and subsequently subjected to electrochemical cycling

analysis. The inset shows a magnified view of the spectral region corresponding to the expected S and Ni edge positions. **(b)** Integrated peak areas corresponding to sulfur (350-480 channels) and nickel (500-625 channels) for the respective Ni implantation fluences.

Tab. 1 summarizes, for Ni fluences of 0 (unirradiated), $1 \times 10^{12}$, $1 \times 10^{13}$, and $5 \times 10^{13}$ cm$^{-2}$, the ratios of integrated areas: Ni/Ni$_{pristine}$, S/S$_{pristine}$, and S/Ni. For simplicity, the influence of electronic artefacts (pile-up effects) was neglected in the integration of the S and Ni regions. As shown, the areas associated with both Ni and S increase with increasing Ni fluence (with the exception of $1 \times 10^{13}$ cm$^{-2}$), while the S/Ni ratio decreases, indicating that the relative amount of sulfur in the foil decreases with respect to nickel.

Assuming that sulfide species are formed in the modified PE separators during H-cell cycling (including in the pristine foil, where trace Ni contamination from the environment is likely present), and considering that the S/Ni ratio for common nickel sulfides is always $\leq 2$, it becomes evident that most of the detected sulfur is not bound to Ni (possibly with the exception of the $1 \times 10^{13}$ cm$^{-2}$ sample). Instead, the sulfur is present in another form (e.g., as dioxide $SO_2$) or is physically trapped within the PE structure (e.g., in nanopores), where it remains even after rinsing the separators following electrochemical cycling.

Tab. 1. Ratios of the integrated areas Ni/Ni$_{pristine}$, S/S$_{pristine}$, and S/Ni for PE samples implanted with 1.1 MeV Ni ions at fluences of 0, $1 \times 10^{12}$, $1 \times 10^{13}$, and $5 \times 10^{13}$ cm$^{-2}$.

| sample / fluence | pristine | 1E12 cm$^{-2}$ | 1E13 cm$^{-2}$ | 5E13 cm$^{-2}$ |
|---|---|---|---|---|
| Ni/Ni$_{pristine}$ | 1 | 1.39 | 1.67 | 3.06 |
| S/S$_{pristine}$ | 1 | 1.31 | 0.74 | 1.73 |
| S/Ni | 5.37 | 5.08 | 2.40 | 3.06 |

For PE separators coated with Ni layers prepared via IBS, the situation is more complex. Fig. 4 present RBS spectra with simulations (a, c, e) and the corresponding depth profiles (b, d, f) for samples with $T_{Ni}$ = 20, 40, and 60 minutes, followed by the electrochemical cycling in the H-cell. As evident, the spectra display, in addition to the characteristic carbon (~170 channels) and fluorine (~320 channels) edges from the PE substrate, prominent peaks corresponding to the dopants S (~470 channels) and Ni (~600 channels), as well as O (~260 channels) arising from oxidation of Ni and S elements within the layer.

The derived depth profiles of both the dopants and the PE substrate components in the modified layer are dependent on the deposited Ni thickness. Due to the porous nature of the separators, Ni and S (and, through their oxidation, O) penetrate beyond the surface coating into deeper regions of PE. This leads to increased intermixing of the dopants with the polymer, which becomes more pronounced with thicker Ni layers. This effect is particularly evident in the sample with the thickest Ni layer (see Fig. 4f).

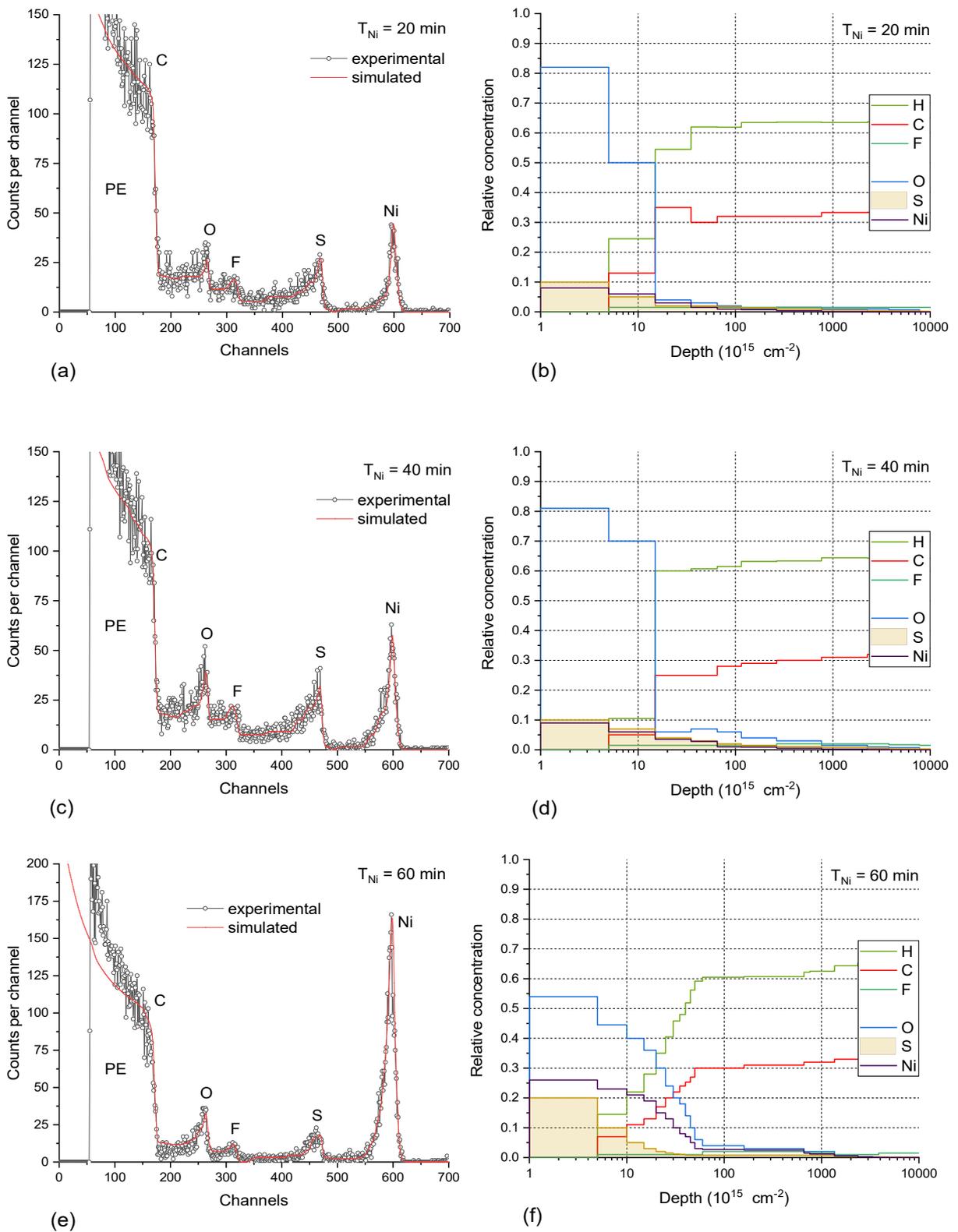

**Fig. 4.** RBS spectra with SIMNRA simulations for PE samples prepared by IBS with Ni deposition times of $T_{Ni}$ = 20 min **(a)**, 40 min **(c)**, and 60 min **(e)**, followed by electrochemical cycling in the H-cell. Corresponding depth profiles (in units of $10^{15}$ cm$^{-2}$) of the PE elements

H, C, and F, and of the dopants Ni, S, and O for Ni deposition times of $T_{Ni}$ = 20 min **(b)**, 40 min **(d)**, and 60 min **(f)**.

Tab. 2 summarizes the amounts of the dopants (Ni, S, O) and the polymer constituent elements (C, H, F) in the modified layer with a total thickness of 2.3 x $10^{19}$ atoms/cm$^2$, corresponding to the observed penetration depth of the in-diffused dopants, for all deposition times $T_{Ni}$. The reported amounts (in units of $10^{15}$ atoms/cm$^2$) were derived from the depth profiles of the elements obtained from detailed RBS simulations. As evident, the content of the dopants changes significantly with the deposited Ni thickness, whereas the quantities of the polymer constituent elements remain approximately constant.

Tab. 2. Amounts of dopants (Ni, S, O) and constituent elements of PE (C, H, F) in the modified layer (with a total thickness of 2.3 x $10^{19}$ atoms/cm$^2$) for Ni deposition times $T_{Ni}$.

| Sample ($T_{Ni}$) | Amount of elements (in 1E15 at/cm$^2$) in a Ni/PE layer with a thickness of 2.3 x E19/cm$^2$ | | | | | |
|---|---|---|---|---|---|---|
| | coating | testing | oxidation | fluorinated polyethylene | | |
| | Ni | S | O | C | H | F |
| 20 min | 8.45 | 46.80 | 78.80 | 7712 | 14719 | 341 |
| 40 min | 13.24 | 61.25 | 103.85 | 7685 | 14534 | 355 |
| 60 min | 41.05 | 27.75 | 65.63 | 7714 | 14579 | 332 |

This trend is further illustrated in Tab. 3, which lists the ratios of elemental contents: Ni/Ni$_{20}$, S/S$_{20}$, Ni/S, Ni/O, S/O, C/H, and C/F (where the subscript 20 refers to the sample with $T_{Ni}$ = 20 min). While the ratios of dopants vary significantly with deposition time, the ratios of the PE elements remain nearly unchanged (with differences < 1% for C/H and < 5% for C/F). Particularly noteworthy are the relative ratios of the dopants. Assuming that NiS compounds are formed and that both elements undergo oxidation in the H-cell, and considering that the most stable nickel sulfide is NiS [27], the most stable nickel oxide is NiO [28], and the most stable and dominant sulfur oxide is SO$_2$ [29], the measured amounts of dopants allow one to unambiguously determine the ratios of Ni/S, Ni/O, and S/O$_2$ (see Tab. 4).

Tab. 3 Ratios of elemental contents for Ni-coated PE samples: Ni/Ni$_{20}$, S/S$_{20}$, Ni/S, Ni/O, S/O, C/H, and C/F in the layer with a total thickness of 2.3 x $10^{19}$ at/cm$^2$.

| Sample ($T_{Ni}$) | Ni/Ni$_{20}$ | S/S$_{20}$ | Ni/S | Ni/O | S/O | C/H | C/F |
|---|---|---|---|---|---|---|---|
| 20 min | 1 | 1 | 0.181 | 0.107 | 0.594 | 0.524 | 22.20 |
| 40 min | 1.567 | 1.309 | 0.216 | 0.127 | 0.590 | 0.529 | 22.64 |
| 60 min | 4.858 | 0.593 | 1.479 | 0.625 | 0.423 | 0.529 | 23.23 |

Analysis of ratios indicates that, for the '20 min' sample, NiS consists of more than 99% of the deposited Ni and 16.6% of S, NiO comprises only 1% Ni and 1% O, and SO$_2$ contains 83.4% S and 99% O. For the '40 min' sample, the distribution differs: NiS is composed of 80.3% Ni and 17.4% S, NiO contains 19.7% Ni and 2.5% O, and SO$_2$ comprises 82.6% S and 97.5% O. The '60 min' sample exhibits a markedly different dopant distribution: NiS consists of 25% Ni and 37.2% S, NiO contains 75% Ni and 46.7% O, and SO$_2$ comprises 62.8% S and 53.1% O.

Tab. 4 Quantitative distribution of dopants forming NiS, NiO, and SO$_2$ in the layer with a total thickness of 2.3 x 10$^{19}$ atoms/cm$^2$.

| Sample (T$_{Ni}$) | Quantitative distribution of dopants (in 1E15 at./cm$^2$) | | |
|---|---|---|---|
| | Ni:S | Ni:O | S:O$_2$ |
| 20 min | 7.75 : 7.75 | 0.7 : 0.7 | 39.05 : 78.1 |
| 40 min | 10.63 : 10.63 | 2.61 : 2.61 | 50.62 : 101.24 |
| 60 min | 10.31 : 10.31 | 30.74 : 30.74 | 17.44 : 34.88 |

Fig. 5 illustrates the dependence of the amounts of quantitatively formed composites on the deposited sample as a function of Ni deposition time (T$_{Ni}$). As evident, the variation of NiS, NiO, and SO$_2$ with T$_{Ni}$ differs for each compound. The NiS sulfide initially increases by 37% between 20 and 40 minutes and then remains nearly constant at T$_{Ni}$ = 60 min. The amount of NiO is very low at T$_{Ni}$ = 20 min, rises sharply (threefold, though still the lowest) at T$_{Ni}$ = 40 min, and becomes the dominant species at T$_{Ni}$ = 60 min, with an increase of almost twelvefold. The SO$_2$ dioxide is the most abundant compound at T$_{Ni}$ = 20 min; its amount further increases by 30% at T$_{Ni}$ = 40 min, but surprisingly drops sharply by 65% at T$_{Ni}$ = 60 min, falling below the NiO content.

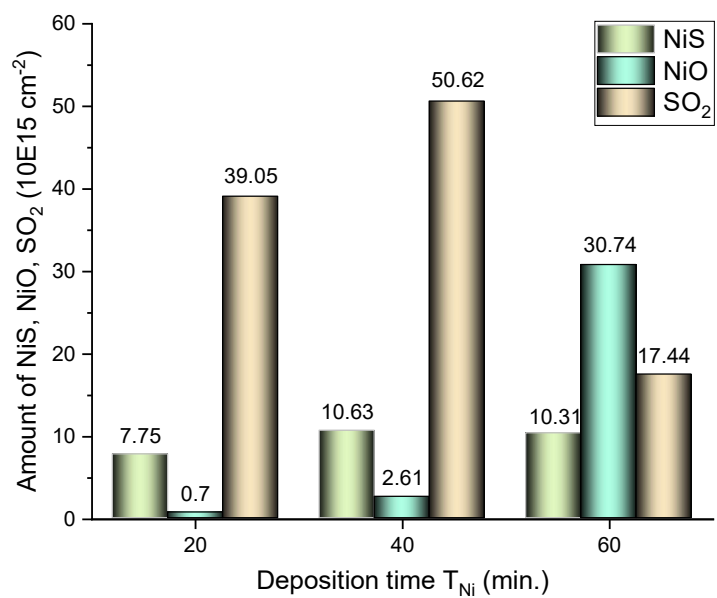

**Fig. 5:** Dependence of the quantitative amounts of the sulfide NiS, and oxides NiO and SO$_2$ on the Ni deposition time (T$_{Ni}$).

From this analysis, it is possible to infer how dopant synthesis and distribution into sulfides and oxides occur in the H-cell. For the sample with T$_{Ni}$ = 20 min, nickel preferentially reacts with sulfur to form stable NiS, producing approximately 11 times less NiO (the production ratio of NiS to NiO depends on specific conditions, but the preferential formation of NiS is desirable [30], as observed in catalytic reactions and attributed to the high activity of the Ni-S interface [31, 32]). The excess sulfur, which is five times greater than Ni, then forms SO$_2$ with oxygen, yielding about 4.6 times more SO$_2$ than the combined NiS + NiO. It should be noted that sulfur oxidation is significantly more exothermic and thermodynamically favorable than the formation of NiS.

For the sample with a higher deposited Ni content ($T_{Ni}$ = 40 min), all three compounds – NiS, NiO, and $SO_2$ increases, while the NiS/NiO ratio decreases markedly to 4.01, and $SO_2$/(NiS + NiO) drops to 3.8, indicating a shift in the dynamics of composite formation. The 60 min sample exhibits a surprising sharp increase in NiO (NiS/NiO = 0.34) accompanied by a pronounced decrease in $SO_2$ [$SO_2$/(NiS + NiO) = 0.42].

These changes could be attributed to oxidation of Ni and S within the sample, with sulfur rapidly diffusing into the Ni layer [33]. Oxidation occurs not only at the Ni surface but also, due to the porous fibrous structure of PE [34] and the permeation of oxygen through it [35], at the interface between the separator and the Ni layer. As a result, NiO forms on the top and, partially, on the bottom side of the Ni layer, with the amount of NiO expected to increase for thicker layers.

It is also important to note that under oxidizing conditions (e.g., exposure to air), NiS can undergo oxidation and spontaneously convert into nickel oxide/oxyhydroxide, NiO(OH) [36]. This process involves the loss of sulfur and incorporation of oxygen into the nickel compound. Furthermore, NiS can also be transformed into NiO during electrochemical testing [37], when NiS reacts with hydroxide ions ($OH^-$) to form nickel hydroxide, $Ni(OH)_2$, which is subsequently converted into NiO or nickel oxyhydroxide, NiOOH.

The observed processes and their dynamics are assumed to be responsible for the changes in the NiS/NiO ratios. Of particular interest is the pronounced decrease in $SO_2$ for the sample with $T_{Ni}$ = 60 min, which is likely related to the increased NiO layer. This layer forms a barrier on the surface, hindering the diffusion of O and S into the deposited Ni layer (diffusion of O and S in NiO is generally much slower than through pure Ni). As a result, the formation of $SO_2$ dioxide, anchored within the deposited layer, becomes more difficult, and the total amount formed is lower than in samples with thinner Ni layers.

The results discussed above indicate that Ni implantation and thin Ni coatings alone are not sufficiently effective at suppressing polysulfide migration. In the next step, a set of modified samples was prepared with LAGP and Ni or Co coatings (deposited at a constant rate for 1 hour to ensure the formation of homogeneous films). The SEM micrographs, shown in Fig. 6, reveal that the surface morphology of these samples still remains fibrous (similar to Fig. 2c) without clustering or agglomeration of nanoparticles.

The electrochemical properties of the pristine and modified PE separators, as well as the modified LAGP samples, were investigated using an Autolab PGSTAT 204 electrochemical workstation. Cyclic voltammetry (CV) was performed within a potential range of ± 1 V. All measurements were carried out in a 1 M LiTFSI solution in a DOL/DME (1:1 v/v) electrolyte. Sulfur electrodes were cut into 25 x 25 mm squares. A copper foil, double-sided coated with graphite, was used as the anode. Graphite is a well-established anode material in lithium-ion batteries and has strong potential as a multifunctional material in lithium-sulfur battery systems.

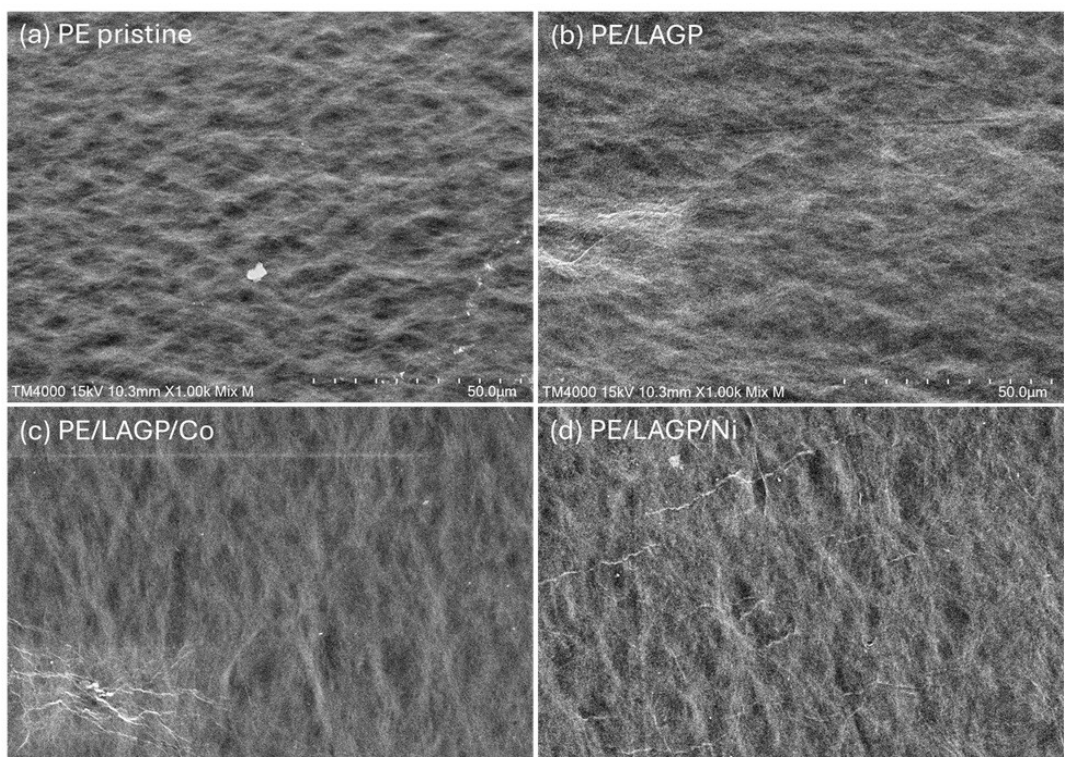

**Fig. 6.** SEM micrographs of the PE separator with a LAGP layer and thin Ni or Co coatings: **(a)** pristine PE, **(b)** PE with LAGP, **(c)** PE with LAGP and Co, **(d)** PE with LAGP and Ni.

Staircase voltammetry (SV) measurements were carried out to assess the influence of thin-film coatings on the electrochemical behavior and cycling stability of the Li-S cells. The resulting SV curves, shown in Fig. 7, represent the averaged response of five consecutive cycles for each sample, ensuring data reliability and reproducibility.

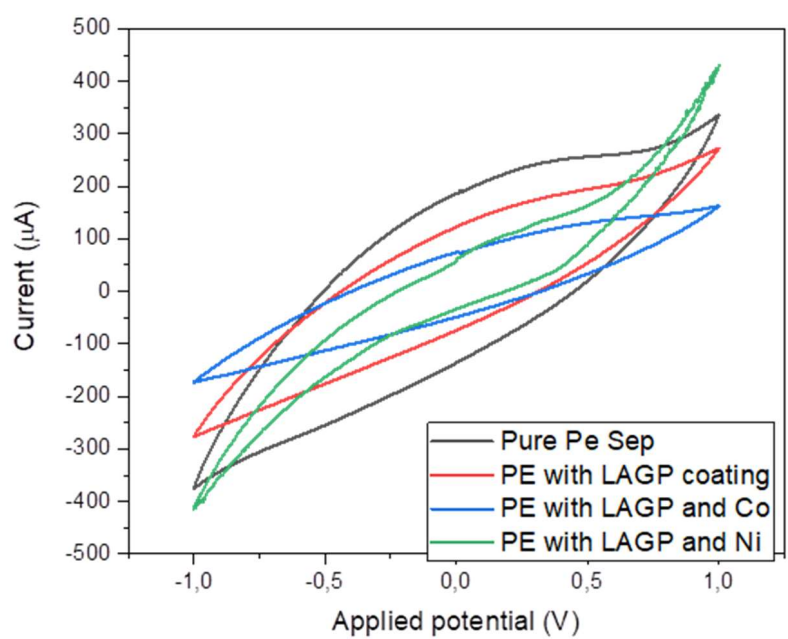

**Fig. 7.** Staircase voltammetry obtained through the PE separator used either as a pristine foil or modified with thin LAGP, LAGP/Ni, and LAGP/Co coatings. SV measurements were performed over five consecutive cycles to assess the influence of the thin-film coatings on the electrochemical response and cycling stability.

The pristine PE separator exhibits pronounced redox peaks and a steep slope, characteristic of intense polysulfide shuttle activity and limited electrochemical stability. In contrast, the introduction of a LAGP layer significantly improves the electrochemical response, as evidenced by reduced current intensity and a narrower hysteresis between the anodic and cathodic branches. This behavior indicates suppressed shuttle reactions and enhanced ionic selectivity, which can be attributed to the high Li-ion conductivity and polysulfide-blocking properties of the LAGP ceramic network. Further modification with a cobalt layer (LAGP/Co) results in an even smoother voltammetry profile with lower current response, demonstrating effective suppression of polysulfide migration and more controlled redox kinetics. The synergistic interaction between the LAGP ionic conductor and the Co layer likely promotes localized catalytic conversion of polysulfides at the cathode while restricting their diffusion through the separator. In contrast, the LAGP/Ni separator exhibits signs of instability, with broader and asymmetric voltammetric curves indicative of parasitic reactions or partial degradation during cycling, consistent with the limited long-term stability of Ni coatings observed previously. Overall, the LAGP/Co configuration demonstrates the most favorable electrochemical performance, effectively combining high ionic conductivity, interfacial stability, and polysulfide shuttle suppression. This design significantly mitigates polysulfide migration, thereby enhancing the cycling efficiency and lifespan of Li-S batteries.

Clear evidence of polysulfide suppression was observed in the H-cell equipped with the carbon electrode. The solvent color, which would normally turn yellow due to dissolved sulfur species, varied depending on the applied coating (Fig. 8). With the pristine PE separator, the LiTFSI-DOL-DME solvent became yellowish; with the LAGP coating, it developed a bluish tint; and with the LAGP/Ni or LAGP/Co coatings, the solution remained transparent. This behavior indicates effective confinement of polysulfides and their inhibited migration through the separator, consistent with the voltammetry results.

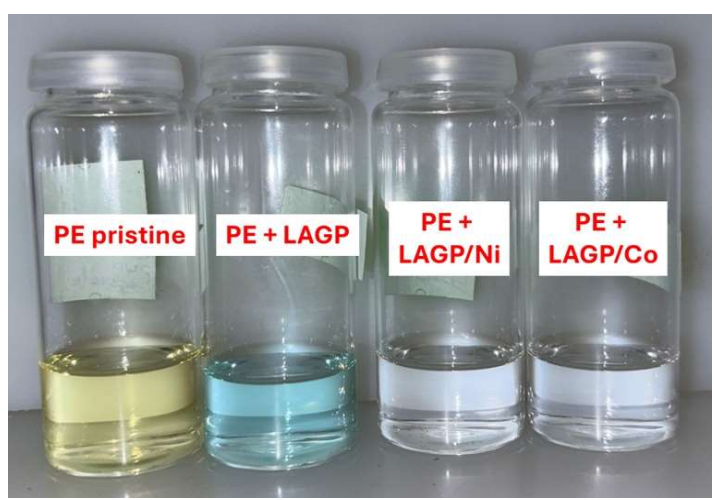

**Fig. 8** Example of solvent color change depending on the PE separator coating. The solvent used for PE pristine is yellowish as it contains sulfur, the solvents for PE/LAGP/Ni or PE/LAGP/Co are clear (as the sulfur content is low).

Additional RBS analyses were carried out to quantify the amount and depth distribution of sulfur trapped in the solid electrolytes after cycling. RBS was also employed to determine the residual sulfur content in the solvents used in the H-cell equipped with PE separators coated with LAGP or LAGP combined with transition metals (pristine PE, LAGP, LAGP/Ni, and LAGP/Co). For this purpose, fresh PE served as the substrate, onto which a small volume (100 µl) of each solvent was deposited. The solvents were applied over a 1 cm$^2$ area of the PE foil and allowed to dry for 24 hours in a controlled environment (vacuum desiccator). Both sample types – PE coated with LAGP-based layers (LAGP, LAGP/Ni, LAGP/Co), and PE substrates containing dried solvent residues – were subsequently analyzed by RBS using a 2 MeV alpha-particle probing beam.

Fig. 9a shows the sulfur depth profiles in the modified separators, evaluated using the SIMNRA code. The simulations reveal a significant difference in the sulfur depth distributions between the pristine PE (or PE coated only with LAGP) and PE coated with LAGP/Ni or LAGP/Co. As evident from the data, the areal density of sulfur in the (0 – 16000) x 10$^{15}$ cm$^{-2}$ subsurface region is very low for pristine PE and for PE with LAGP (on the order of 1 x 10$^{15}$ cm$^{-2}$). In contrast, at greater depths, specifically within the (16000 – 22000) x 10$^{15}$ cm$^{-2}$ region, the sulfur areal density increases to 135 x 10$^{15}$ cm$^{-2}$, and in the deeper (22000 – 31000) x 10$^{15}$ cm$^{-2}$ region it reaches a maximum value of 155 x 10$^{15}$ cm$^{-2}$.

The table-shaped sulfur profile with a depleted subsurface region is attributed to the fibrous morphology of PE. In this case, diffusion is strongly influenced by the fibers and their orientation, and is typically faster in-plane (parallel to the fibers) [38]. Additionally, the surface chemistry of the fibers can affect molecular adhesion and thereby alter diffusion rates. Treatments such as irradiation or the application of coatings can further modify this behavior. This effect is evident in PE with the combined LAGP/Ni or LAGP/Co overlayers, where the sulfur areal density in the subsurface region (comprising LAGP and the ferroic metals) is substantially higher: 75 x 10$^{15}$ cm$^{-2}$ for Ni and 55 x 10$^{15}$ cm$^{-2}$ for Co coating, respectively.

At intermediate depths (13000 – 16000 x 10$^{15}$ cm$^{-2}$), the areal density decreases sharply to 11 × 10$^{15}$ cm$^{-2}$ for Ni and to approximately 1 × 10$^{15}$ cm$^{-2}$ for Co coatings. At larger depths (13000 – 16000 x 10$^{15}$ cm$^{-2}$), the sulfur areal densities become comparable to those observed for pristine PE (and PE with LAGP). However, in the upper portion of this region (16000 – 22000 x 10$^{15}$ cm$^{-2}$), the areal densities differ slightly: 145 x 10$^{15}$ cm$^{-2}$ for Ni and 115 x 10$^{15}$ cm$^{-2}$ for Co – somewhat lower than the values obtained for pristine PE.

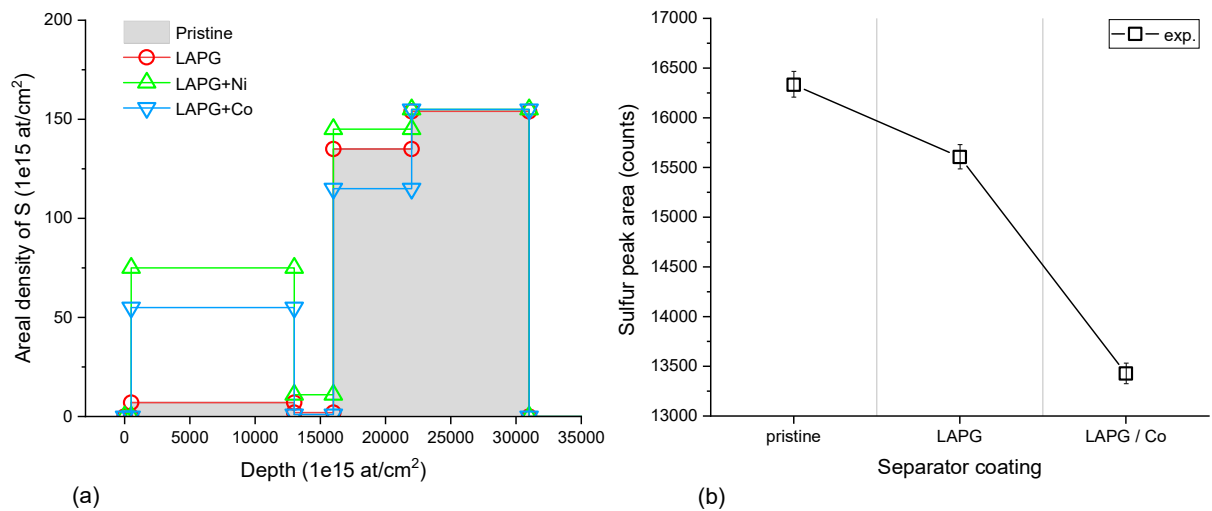

(a)  (b)

**Fig. 9. (a)** Sulfur depth profiles in pristine PE, PE/LAGP, PE/LAGP/Ni, and PE/LAGP/Co separators. **(b)** Amount of sulfur detected by RBS in the solvents (dried on PE) used in the electrochemical cycling for pristine PE, PE with LAGP, and PE with LAGP/Co.

Fig. 9b shows the amount of sulfur detected in the 100 μl of solvent used for the analyzed samples (with the exception of LAGP/Ni). These values are proportional to the total sulfur concentration dissolved in the solvents. It should be noted that lower detected amounts indicate stronger suppression of sulfur diffusion in modified separators, associated with increased sulfide/oxide formation, and vice versa. As shown, the lowest residual sulfur content in the solvent was observed for the sample with the LAGP/Co coating (the LAGP/Ni sample exhibited a similar level, data are not shown here), whereas the highest sulfur content was found for the pristine PE and PE/LAGP samples. This trend is consistent with the sulfur depth profiles presented in Fig. 9a.

Overall, the results in Fig. 9 highlight the role of ferroic layers in inhibiting sulfur transport through ion-conductive PE separators, a function that is crucial for thin solid-state Li-ion batteries employing sulfur-based anodes.

## Conclusions

This pilot study explores the application of low-energy ion beams for the fabrication of ultrathin solid-electrolyte films, with the goal of advancing thin all-solid-state Li-ion batteries employing sulfur cathodes. The results demonstrate effective suppression of the shuttle effect in the presence of the LISICON electrolyte, particularly when ferroic elements (Ni, Co) are incorporated at its surface. Staircase voltammetry performed on PE separators coated with LAGP and Ni (or Co) multilayers reveals diminished redox peak intensities and lower slopes, indicative of enhanced electrochemical stability and improved overall performance. Given that polysulfide suppression remains a central challenge hindering the commercialization of Li-S batteries, these findings point to a promising strategy for strengthening their energy-storage capability. Overall, the combined use of Ni and LAGP coatings emerges as a compelling avenue for future investigation.

## Acknowledgements

This work was supported by the Ministry of Education, Youth and Sports (MEYS) CR under the project OP JAK CZ.02.01.01/00/22_008/0004591. The experiment was carried out at the CANAM infrastructure (project no. LM2015056).